\documentclass[acus]{jacow}
\usepackage{graphicx}  
\usepackage{subfigure} 
\usepackage{bm}        
\usepackage{amsmath}   
\usepackage{cancel} 
\usepackage{verbatim} 
\usepackage{hyperref}
\usepackage{cite}

\begin{document}

\title{Space-Charge Simulation of Integrable Rapid Cycling Synchrotron}
\author{J. Eldred and A. Valishev, FNAL, Batavia, Illinois 60510 USA}

\maketitle

\begin{abstract}

Integrable optics is an innovation in particle accelerator design that enables strong nonlinear focusing without generating parametric resonances. We use the Synergia tracking code to investigate the application of integrable optics to high-intensity hadron rings. We consider an integrable rapid-cycling synchrotron (iRCS) designed to replace the Fermilab Booster. We find that incorporating integrability into the design suppresses the beam halo generated by a mismatched KV beam. Our iRCS design includes other features of modern ring design such as low momentum compaction factor and harmonically canceling sextupoles. Experimental tests of high-intensity beams in integrable lattices will take place over the next several years at the Fermilab Integrable Optics Test Accelerator (IOTA) and the University of Maryland Electron Ring (UMER).

\end{abstract}

\section*{Background}

Integrable optics is a development in particle accelerator technology that enables strong nonlinear focusing without generating new parametric resonances~\cite{Danilov}. A promising application of integrable optics is in high-intensity rings, where nonlinearity is known to suppress halo formation~\cite{Sonnad,WebbArXiV} and enhance Landau damping of charge-dominated collective instabilities~\cite{Macridin}

The efficacy of an accelerator design incorporating integrable optics will undergo comprehensive experimental tests at the Fermilab Integrable Optics Test Accelerator (IOTA)~\cite{Antipov} and the University of Maryland Electron Ring (UMER)~\cite{Ruisard} over the next several years. 

At Fermilab, a core research priority is to improve the proton beam power for the high-energy fixed target program~\cite{Prebys}. The LBNF/DUNE high-energy neutrino program~\cite{DUNE} in particular requires at least 900 kt$\cdot$MW$\cdot$year neutrino exposure for a comprehensive measurement of the CP-violating phase~\cite{Prebys}.

The Proton Improvement Plan II (PIP-II) will replace the 400 MeV linac with a new 800 MeV linac that will increase the 120 GeV proton power of the Fermilab complex to 1.2 MW~\cite{PIP2}. But to achieve a high-energy proton power significantly beyond 1.2 MW, it will be necessary to replace the Fermilab Booster with a modern RCS~\cite{Prebys,EldredNAPAC}.

In this paper, we present our space-charge simulations of an integrable rapid-cycling synchotron (iRCS) design to investigate the performance of integrable optics in this context.

\section*{iRCS Example Lattice}

In \cite{Danilov} a procedure for integrable accelerator design is derived based on an alternating sequence of linear and nonlinear sections.The linear-sections, known as T-inserts, are arc sections with $\pi$-integer betatron phase-advance in the horizontal and vertical plane. The lattice should be dispersion-free in the nonlinear section, and the horizontal and vertical beta functions should be matched. A special nonlinear elliptical magnet is matched to the beta functions to provide the nonlinear focusing. The manipulation of the beta functions and phase-advances removes the time-dependence of the nonlinear kick so as to avoid introducing parametric resonances.

In \cite{EldredNAPAC} we introduced a specific design of an iRCS that meets the essential single-particle requirements of an iRCS - periodicity, bounded beta function, low momentum compaction factor, long dispersion-free drifts, and Danilov-Nagaitsev integral accelerator design.

Modern RCS design also includes sextupoles for chromaticity correction. Furthermore integrability requires that the horizontal and vertical chromaticity be matched~\cite{Webb,Cook}. Using weak sextupoles only to match the horizontal and vertical chromaticity, the chromaticity is -33. Strong sextupoles can be used to correct this chromaticity to -7.7 or any value in between.

To maintain integrability, sextupole magnets should also be located so that their effect cancels harmonically within the T-insert arc. In this work, we carefully managed the phase advances between the sextupoles to cancel the third-order harmonic resonance driving terms. In general, these third-order resonance driving terms are governed by
\begin{align} \nonumber
G_{3,0,l} & \propto \int_{0}^{L} \beta_{x}^{3/2}(s) S(s) e^{j[3\psi_{x}(s)]} ds \\
G_{2,\pm 1,l} & \propto \int_{0}^{L} \beta_{x}^{1/2}(s) \beta_{y} (s) S(s) e^{j[\psi_{x}(s)\pm 2\psi_{y}(s)]} ds 
\end{align}
as given in \cite{SYBook}.

The lattice can be optimized to anticipate space-charge tune depression - the phase advances between the nonlinear inserts and between sextupoles are correct when the beam intensity matches its design value.

\begin{figure}[h]
\begin{centering}
  \includegraphics[height=200pt, width=220pt]{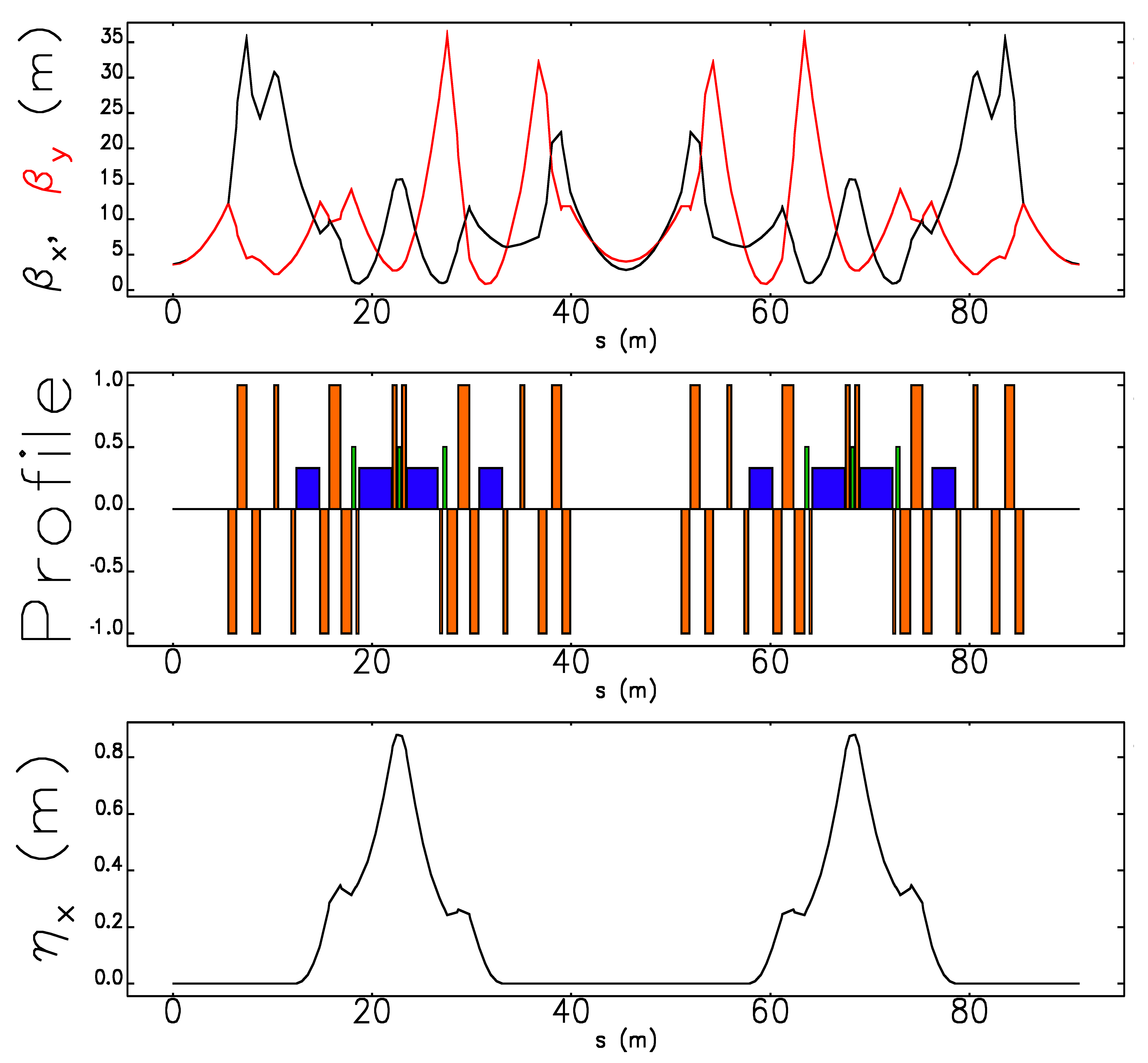}
  \caption{Twiss parameters for one of the six periodic cells. (top) Horizontal and vertical beta functions shown in black and red, respectively. (middle) Location and length of magnetic lattice, elements where dipoles are shown as short blue rectangles, quadrupoles as tall orange rectangles, and sextupoles as green rectangles. (bottom) Linear dispersion function.}
  \label{Lattice}
\end{centering}
\end{figure}

Figure~\ref{Lattice} shows this version of the iRCS lattice and Table~\ref{Param} shows the key parameters of this lattice.

\begin{table}[htp]
\centering
\caption{Parameters of iRCS Lattice}
\begin{tabular}{| l | c |}
\hline
Parameter & Value \\
\hline
Circumference & 546 m \\
Periodicity & 6 \\
Bend Radius & 15.6 m \\
Max Beta Function  & 35 m \\
Max Dispersion & 0.8 m \\
\hline
Insertion Length & 11.2 m \\
Phase-advance over insert & 0.3 $\pi$ \\
Betatron Tune & 16.8 \\
Matched Chromaticity & -33 \\
Corrected Chromaticity & -7.7 \\
Momentum Compaction & 3.8 $ \times 10^{-3}$ \\
\hline
\end{tabular}
\label{Param}
\end{table}

\section*{Space-charge Simulation Framework}

This example iRCS lattice is designed to serve as a versatile platform for a variety of multiparticle simulation experiments. We use Synergia to calculate how the space-charge forces interact with the nonlinear optics. Synergia is a Python-based parallel code for multiparticle tracking with imported CHEF functionality~\cite{Amundson}. There is an ongoing effort to simulate the IOTA lattice using Synergia~\cite{Bruhwiler} and to benchmark it against IMPACT-Z. 

Space-charge forces were calculated using the Synergia 2D-Hockney solver, a 2D particle-in-cell method described in \cite{Holmes} and \cite{Galambos}. For this work we used 8 steps per element, 32 $\times$ 32 space-charge grid, map order 6, and $10^{5}$ macroparticles.

\section*{Halo from Mismatched KV Beam}

An important source of transverse beam halo is described by the particle-core model~\cite{Wangler}. In the particle-core model, the breathing mode oscillations of a uniform-density beam core can drive particles into the halo. We simulate a mismatched KV beam in our lattice in order to generate a source of halo and evaluate the impact of nonlinear integrable optics.

We compare two similar versions of the lattice, which we refer to as the conventional case and the integral case. In the conventional case, the nonlinear insert is deactivated and there are strong sextupoles for chromaticity correction. In the integrable case, the nonlinear insert is activated and there are weak sextupoles only for matching the horizontal and vertical chromaticity. The two cases represent two different approaches to addressing collective instabilities, but those instabilities are not simulated in this work.

In the integral case we have a phase advance through the nonlinear insert of $Q_{0}=0.3$, a strength parameter of $t=0.15$, and elliptic parameter $c=0.16$ m$^{1/2}$ (normalized coordinates). The corresponding nonlinear small-amplitude tune shifts are $\Delta Q_{x} = Q_{0} (\sqrt{1+2t}-1) = 0.04$ and $\Delta Q_{y} = Q_{0} (\sqrt{1-2t}-1) = -0.05$ (see \cite{Danilov,Nagaitsev}). 

For this test, the beam energy is 0.8 GeV and the initial 95\% normalized emittance is 20 mm mrad (or normalized rms emittance of 5 mm mrad). The KV space-charge tune-shift is 0.05 corresponding to a ring intensity of $12 \times 10^{12}$ protons unbunched beam.

The KV beam is generated with a uniform sampling of phase-space coordinates with a Hamiltonitan value $H_{0}$. The initial distribution for the integral case will not be the same as the conventional case, because the KV distribution for the integral case follows the nonlinear equipotential contours. A 5\% mismatch is introduced by scaling the initial $y$-coordinate by 0.95 and the initial $x$-coordinate by 1.05 while preserving the initial $x^{\prime}$ and $y^{\prime}$ coordinates. There is no initial momentum spread.

\begin{figure}[htb]
\begin{centering}
  \includegraphics[height=100pt, width=220pt]{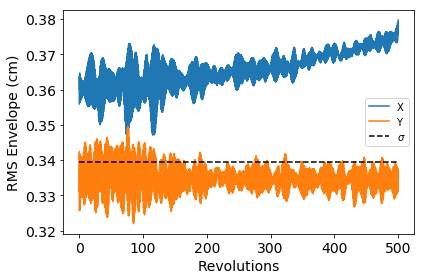}
  \includegraphics[height=100pt, width=220pt]{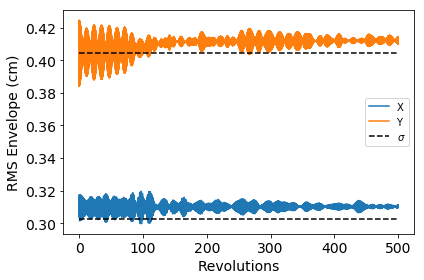}
  \caption{The rms beam size in the horizontal (blue) and vertical (orange) for conventional (top) and integrable (bottom) case. Black dashed lines indicates nominal beam rms without mismatch.}
  \label{RMS}
\end{centering}
\end{figure}

Figure~\ref{RMS} shows the rms beam size for the conventional and integrable case at the center of the nonlinear insert. In the conventional case, the horizontal rms size grows steadily. In the integrable case, the quadrupole oscillation damps rapidly over the first 150 revolutions. The nonlinear elliptic focusing in the integrable case also causes the equilibrium beam size to be larger in the vertical plane.

\begin{figure}[htb]
\begin{centering}
  \includegraphics[height=120pt, width=220pt]{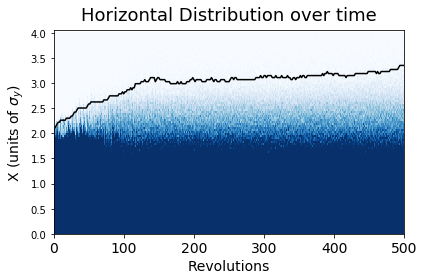}
  \includegraphics[height=120pt, width=220pt]{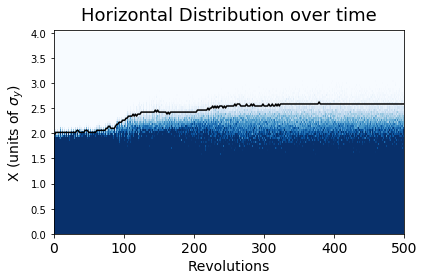}
  \caption{Horizontal particle distribution over time for conventional (top) and integrable (bottom) case. The color axis is scaled to express the variation in the halo density instead of the full density range. The black line indicates the 99.9th percentile of the beam (100 macroparticles).}
  \label{Halo}
\end{centering}
\end{figure}

Figure~\ref{Halo} shows the horizontal particle distribution over the same time scale. The halo formation is continuously driven in the conventional case and suppressed in the integrable case.

\begin{figure}[htb]
\begin{centering}
  \includegraphics[scale=0.18]{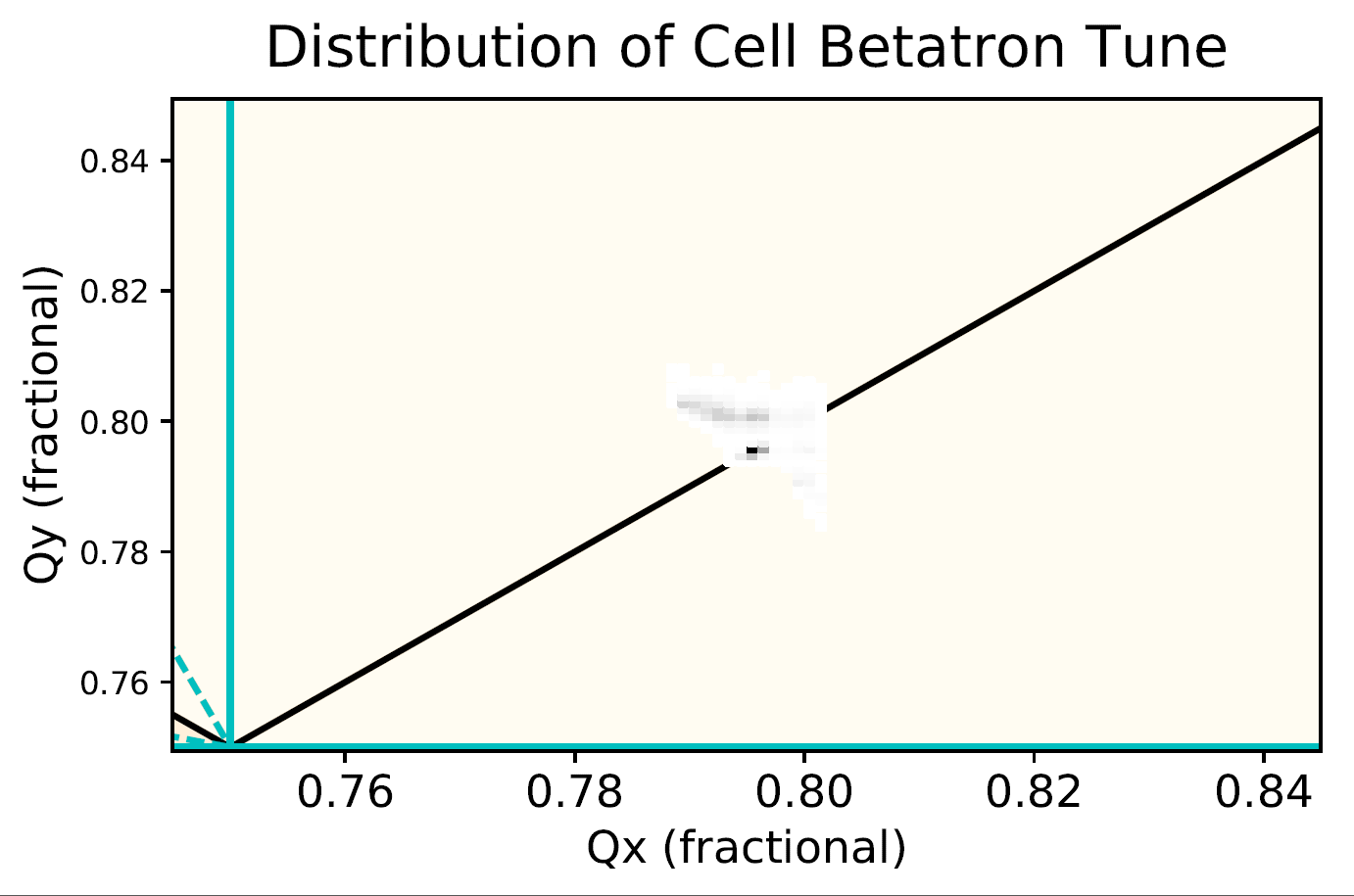}
  \includegraphics[scale=0.18]{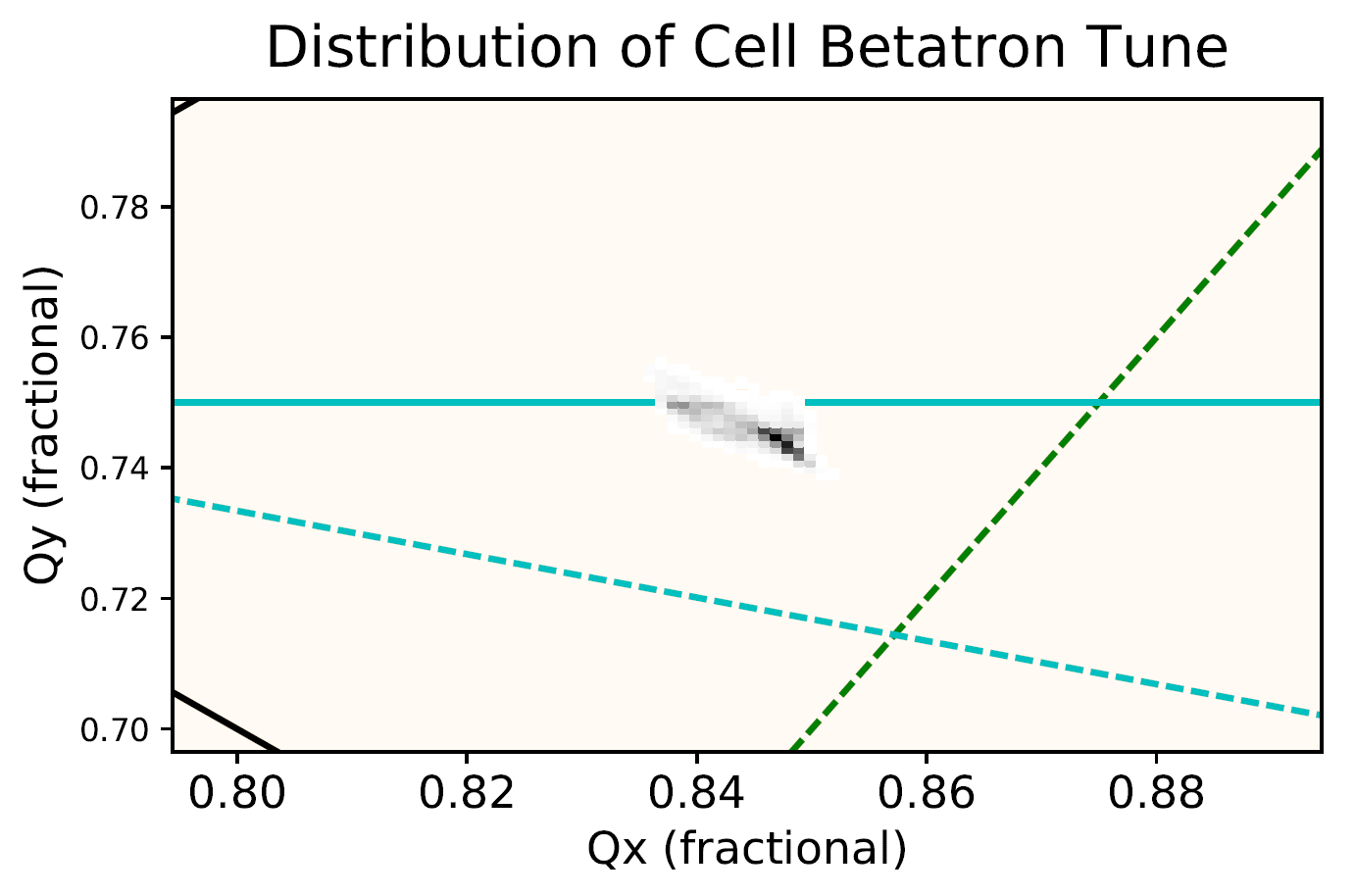}
  \caption{Betatron tune diagram across one periodic cell for conventional (top) and integrable (bottom). Darker points indicate greater density of particles. The scale in the horizontal and vertical axis is the same for both plots.}
  \label{Tune}
\end{centering}
\end{figure}

Figure~\ref{Tune} shows the tune distribution across one periodic cell (1/6 of iRCS ring). The quadrupole component of the nonlinear insert causes the integral case to have an off-diagonal tune shift relative to the conventional case. We see for this particular value of the nonlinear strength parameter $t$ the integrable case lies directly on the fourth order structural resonance line. As expected, the nonlinear insert does not drive the fourth-order resonance even though it has a significant octupole component.

\section*{Conclusions and Future Work}

We have updated our iRCS lattice design to include harmonically canceling sextupoles and greater phase advance through the nonlinear section. We have begun to perform a variety of space-charge simulation experiments with this lattice using Synergia. Our first major experiment is to study the transverse dynamics of mismatched KV distributions. In this work we confirm that our design incorporating integrable optics provides superior suppression of beam halo even while spanning a fourth-order resonance line.

Our next step is to simulate a beam with waterbag distribution at the beam intensities required for the RCS to support multi-MW operation of the Main Injector. Subsequently we should extend these 2D space-charge simulations with coasting beam to a full 3D space-charge simulation with bunched beam.

There should be a more optimal iRCS lattice that could have higher periodicity, greater compactness and/or the elimination of sextupoles. We also plan to generate a version of an iRCS lattice which breaks the periodicity with random quadrupole errors and investigate the impact of integrable optics on the dynamics aperture.

\section*{Acknowledgments}

We would like to thank the entire Radiasoft team for helping us setup our Synergia environment and for providing valuable feedback throughout our research process.

Operated by Fermi Research Alliance, LLC under Contract No. DE-AC02-07CH11359 with the United States Department of Energy.


\begin{thebibliography}{99}
  
  \bibitem{Danilov} V.~Danilov and S.~Nagaitsev, Phys. Rev. ST Accel. Beams, {\bf 13}, 084002 (2010) [\url{https://journals.aps.org/prab/abstract/10.1103/PhysRevSTAB.13.084002}].
  
  \bibitem{Sonnad} K.~G.~Sonnad and J.~R.~Cary, Phys. Rev. ST Accel. Beams, {\bf 8}, 064202 (2005) [\url{https://journals.aps.org/prab/abstract/10.1103/PhysRevSTAB.8.064202}]
  
  \bibitem{WebbArXiV} S.~Webb, D.~Bruhwiler, D.~Abell et al.,  	arXiv:1205.7083 [\url{https://arxiv.org/abs/1205.7083}]
  
  \bibitem{Macridin} A.~Macridin, A.~Burov, E.~Stern et al., Phys. Rev. ST Accel. Beams, {\bf 18}, 074401 (2015) [\url{https://journals.aps.org/prab/abstract/10.1103/PhysRevSTAB.18.074401}]
  
  \bibitem{Antipov} S.~Antipov, D.~Broemmelsiek, D.~Bruhwiler et al., JINST, {\bf 12}, T03002 (2017)  [\url{http://iopscience.iop.org/article/10.1088/1748-0221/12/03/T03002/meta}]
  
  \bibitem{Ruisard} K.~Ruisard, B.~Beaudoin, I.~Haber et al., Proc IPAC2015 [\url{http://accelconf.web.cern.ch/AccelConf/IPAC2015/papers/mopma046.pdf}]
  
  \bibitem{Prebys} E.~J.~Prebys, P.~Adamson, S.~Childress et al., Proc. of IPAC2016, [\url{http://accelconf.web.cern.ch/AccelConf/ipac2016/papers/tuoaa03.pdf}].
  
  \bibitem{DUNE} R.~Acciarri {\it et al.} [DUNE Collaboration], Long-Baseline Neutrino Facility (LBNF) and Deep Underground Neutrino Experiment (DUNE) Conceptual Design Report Volume 1: The LBNF and DUNE Projects, Fermilab 2015.
  
  \bibitem{PIP2} P.~Derwent {\it et al.}, Fermilab Report No. Project X-doc-1232, 2013, [\url{http://projectx-docdb.fnal.gov/cgi-bin/RetrieveFile?docid=1232;filename=1.2%20MW%20Report_Rev5.pdf;version=3}].
  
  \bibitem{EldredNAPAC} J.~Eldred and A.~Valishev, Proc. NAPAC2016 \url{http://inspirehep.net/record/1516034}
  
  \bibitem{Webb} S.~Webb, D.~Bruhwiler, A.~Valishev et al., presented at AAC14 [\url{https://indico.fnal.gov/getFile.py/access?contribId=26&sessionId=6&resId=0&materialId=slides&confId=8713}]

  \bibitem{Cook} N.~Cook, S.~Webb, D.~Bruhwiler et al., presented at AAC16 [\url{https://indico.syntek.org/event/4/session/14/contribution/142/material/slides/0.pdf}]
  
  \bibitem{SYBook} S.~Y.~Lee, {\it Accelerator Physics}, 3rd Ed, ISBN: 978-9-8143-7494-1, World Scientific, Singapore, (2012).
  
  \bibitem{Amundson} J.~Amundson, Q.~Lu, and E.~Stern, Proc. PYHPC 2013 [\url{http://lss.fnal.gov/archive/2013/conf/fermilab-conf-13-548-apc-cd.pdf}]
  
  \bibitem{Bruhwiler} N.~Cook, D.~Bruhwiler, C.~Hall et al., these proceedings.
  
  \bibitem{Holmes} J.~A.~Holmes, J.~D.~Galambos, D.~K.~Olsen et al. Proc. EPAC98 [\url{http://accelconf.web.cern.ch/AccelConf/e98/PAPERS/THP24C.PDF}]
  
  \bibitem{Galambos} J.~D.~Galambos, S.~Danilov, D.~Jeon et al., Phys. Rev. ST Accel. Beams, {\bf 3}, 034201 (2000) [\url{https://journals.aps.org/prab/abstract/10.1103/PhysRevSTAB.3.034201}]
  
  \bibitem{Wangler} T.~P.~Wangler, K.~R.~Crandall, R.~Ryne et al., Phys. Rev. ST Accel. Beams, {\bf 1}, 084201 (1998) [\url{https://journals.aps.org/prab/abstract/10.1103/PhysRevSTAB.1.084201}]
  
  \bibitem{Nagaitsev} S.~Nagaitsev, A.~Valishev, V.~Danilov et al., Proc. IPAC2012 [\url{accelconf.web.cern.ch/AccelConf/IPAC2012/papers/tuppc090.pdf}]

  S.~Nagaitsev, A.~Valishev and V.~Danilov, Proc. HB2010 [\url{accelconf.web.cern.ch/AccelConf/HB2010/papers/tho1d01.pdf}]
  


\end{thebibliography}
\end{document}